%% file: 100-Hhh.tex
\documentclass[preprint,preprintnumbers,amsmath,amssymb,floatfix,nofootinbib]{revtex4}
\usepackage{epsf, array, color}
\usepackage{graphicx}
\usepackage{subfigure}
\usepackage{bbold}
\usepackage{amsmath}
\usepackage{mathtools}
\usepackage{dcolumn}

\newcommand{\bea}{\begin{eqnarray}}
\newcommand{\eea}{\end{eqnarray}}

\newcommand{\Tr}{\operatorname{Tr}}

\def\beq{\begin{equation}}
\def\eeq{\end{equation}}
\newcommand{\vev}[1]{\langle {#1} \rangle}

\preprint{SLAC-PUB-16119}

\begin{document}

\title{Exploring Resonant di-Higgs production in  the Higgs Singlet Model}

\author{Chien-Yi~Chen$^{\, a}$, S.~Dawson$^{\, a}$ and I.~M.~Lewis$^{\, a,b}$ }
\affiliation{
\vspace*{.5cm}
  \mbox{$^a$Department of Physics,\\
  Brookhaven National Laboratory, Upton, N.Y., 11973,  U.S.A.}\\
 \mbox{$^b$ SLAC National Accelerator Laboratory,\\
 2575 Sand Hill Rd, Menlo Park, CA, 94025, U.S.A.}
 \vspace*{1cm}}

\date{\today}

\begin{abstract}
We study the enhancement of the di-Higgs production cross section resulting from the resonant decay of a heavy Higgs boson at hadron colliders in a model with a Higgs singlet.
 This enhancement of the double Higgs production 
rate is crucial in understanding the structure of the scalar potential and we determine the maximum allowed
enhancement such that the electroweak minimum is a global minimum. 
The di-Higgs production enhancement can be as large as a factor of $\sim 18 (13)$ for the mass of the heavy Higgs around $270 (420)$ GeV relative to the Standard Model rate at 14 TeV for parameters corresponding to a global electroweak minimum.
\end{abstract}

\maketitle

\section{Introduction}
\input{intro.tex}

\section{Model}
\label{sec:model}
\input{model.tex}

\section{Resonant di-Higgs Production}
\label{sec:results}
\input{di-Higgs.tex}

\section{Experimental and theoretical constraints}
\label{sec:limits}
\input{limits.tex}

\section{Discussion and Conclusions}
\input{conclusions.tex}

\section*{Acknowledgements}
This work is supported by the U.S. Department of Energy under grant
No.~DE-AC02-98CH10886 and contract DE-AC02-76SF00515. 

\appendix
\section{Cubic and quartic couplings}
\input{appendixa.tex}
\section{v=0 solutions}
\input{appendixb.tex}

\newpage
\bibliographystyle{h-physrev}
\bibliography{100-Hhh}

\end{document}

%% file: intro.tex
After the discovery of the Higgs boson, the next task is to determine its couplings to as many Standard Model (SM) 
 particles as possible.  Only by doing so can the true nature of electroweak symmetry breaking be determined.  It is particularly
important to measure the parameters of the scalar potential, which entails measuring double 
Higgs production \cite{Plehn:1996wb,Baur:2002qd,Baglio:2012np}.
In the SM, this rate is small at the LHC \cite{Dittmaier:2011ti,Frederix:2014hta,Dolan:2012rv,Dawson:1998py,deFlorian:2014rta,Grigo:2014jma}, but may be significantly enhanced in models with new physics.  
  One simple extension of
the SM is to add a scalar, $S$, which is a singlet 
under all the gauge 
symmetries \cite{Pruna:2013bma,Barger:2007im,O'Connell:2006wi,Profumo:2014opa}.  
After electroweak symmetry
breaking, $S$ can  mix with the SM Higgs boson, leading to a modification of Higgs couplings to SM 
particles and to the parameters of the scalar potential.  In such models, there can be an enhancement of the di-Higgs rate due to the resonant production of the new scalar \cite{Bowen:2007ia,Barger:2014taa,Dolan:2012ac}. 

Models with a Higgs singlet are highly motivated by Higgs portal 
models \cite{Schabinger:2005ei,Englert:2011yb,Englert:2011aa}
.
  In such models, $S$ is the only particle which couples
to a dark matter sector.  Couplings of the dark matter to the known particles occur only through the mixing of
$S$ with the SM Higgs boson.
If the Higgs singlet model   possesses 
a $Z_2$ symmetry, the scalar singlet itself could be  a dark matter candidate.  Without a $Z_2$ symmetry,  cubic and linear self-coupling terms are allowed in the 
scalar potential and a strong first order electroweak  phase transition is allowed. 
Motivated by the possibility of explaining 
electroweak baryogenesis \cite{Profumo:2007wc,Espinosa:2011ax,Curtin:2014jma},
we examine enhanced double Higgs production
in a model with a scalar singlet and no $Z_2$ symmetry.
 The requirement that the electroweak minimum be
a global minimum provides stringent restrictions on the allowed parameter space.

Attempts to increase the di-Higgs production rate by adding new particles which contribute to double Higgs production 
from gluon fusion have generally not found increases of more than a factor of $2-3$
over the SM rate \cite{Gillioz:2012se,Dawson:2012mk,Chen:2014xwa}.  More successful has been
the study of resonant enhancements, where increases  up to a factor of  $\sim 50$
 relative to the SM prediction  for double Higgs production 
have been found in 2 Higgs doublet models
and the MSSM \cite{Baglio:2014nea,Hespel:2014sla,Bhattacherjee:2014bca,Arhrib:2009hc,Craig:2013hca}.   We determine the maximum allowed enhancement  from
resonant di-Higgs production in the singlet model without a $Z_2$ symmetry \cite{No:2013wsa},  such that the parameters correspond to a
global electroweak minimum \cite{Espinosa:2011ax}.   This case has a number of novel features in comparison
with the well studied $Z_2$ symmetric singlet model \cite{Pruna:2013bma}.

In Section \ref{sec:model}, we review the Higgs singlet model and the minimization of the potential.  Our results for the 
maximum allowed enhancement of the di-Higgs cross section, subject to the restriction that
the electroweak minimum be a global minimum,  are in Section \ref{sec:results}. Experimental 
constraints and theoretical restrictions on the parameters are given in Section \ref{sec:limits}.
We include $2$ appendices:  Appendix A has the complete set of cubic and quartic Higgs self-couplings and Appendix 
B includes a description of the vacuum with $v=0$.

%% file: model.tex
We consider a model containing the SM Higgs doublet, $H$, and an additional
Higgs singlet,  $S$.
The most general scalar potential is,
\begin{equation}
V(H,S) = V_{H}(H) + V_{HS}(H,S) + V_S(S),
\label{vs}
\end{equation}
with
\begin{eqnarray}
V_{H}(H) &=& - \mu^2 \, H^\dagger H + \lambda (H^\dagger H)^2\\
V_{HS}(H,S) &=& \frac{a_1}{2} \, H^\dagger H \, S
   + \frac{a_2}{2} \, H^\dagger H \, S^2\\
V_S(S) &=& b_1 S + \frac{b_2}{2} S^2 + \frac{b_3}{3} S^3 + \frac{b_4}{4} S^4.
\label{potential}
\end{eqnarray}
We do not assume a $Z_2$ symmetry which would prohibit $a_1$, $b_1$ and $b_3$.  The neutral component of the doublet $H$  is  denoted by $\phi_0=(h+v)/\sqrt{2}$, where the vacuum expectation value (vev) is $\langle \phi_0\rangle = {v\over \sqrt{2}}$.  
Similarly, the vev of $S$ is defined as $x$.

The extrema of the potential are
obtained by requiring $\partial V(v,x)/\partial v=0$ and $\partial V(v,x)/\partial x=0$,\footnote{The discussion in this section closely follows that of 
Ref. \cite{Espinosa:2011ax}.}
\begin{eqnarray}
{v\over 2}(-2\mu^2+2\lambda v^2 +a_1 x +a_2 x^2)=0, \label{dVdh}\\
x(b_2+b_3 x +b_4 x^2 +{v^2 \over 2} a_2 )+b_1+{v^2 \over 4} a_1=0.\label{dVds}
\end{eqnarray}

Solving  Eqs.~\ref{dVdh} and \ref{dVds} produce many possible extrema of the potential.  We require that one of these extrema correspond to the electroweak symmetry breaking (EWSB) minimum, $v=v_{EW}=246$~GeV. 
It is important to note that a shift of the singlet field by $S\rightarrow S+\Delta_S$ is just
a redefinition of the parameters of Eq. \ref{potential} and does not change the physics.  Hence, we are free to choose our EWSB minimum as $(v,x)\equiv(v_{EW},0)$, since changing $x$ would correspond to shifting the singlet field.

With this criteria, solving Eqs.~\ref{dVdh} and \ref{dVds} produces,
\begin{eqnarray}
\mu^2=\lambda\,v^2_{EW},\quad\quad b_1=-\frac{v^2_{EW}}{4}a_1.\label{mu2b1}
\end{eqnarray}
Using these solutions, the potential can be written in a more suggestive form, in terms of the 
neutral  component of the Higgs field:
\begin{eqnarray}
V(\phi_0,S)&=&\lambda\left(\phi_0^2-{v^2_{EW}\over 2}\right)^2
+\frac{a_1}{2} (\phi_0^2-{v^2_{EW}\over 2})S
+\frac{a_2}{2}(\phi_0^2-{v^2_{EW}\over 2})S^2\nonumber\\
&&+\frac{1}{4}\left(2b_2+a_2v^2_{EW}\right)S^2+\frac{b_3}{3}S^3+\frac{b_4}{4}S^4,\label{Vnew}
\end{eqnarray}
where an arbitrary constant factor has been dropped.  Then $v =v_{EW}$ and $x =0$ is a minimum by construction. 

\subsection{Scalar Masses and Mixing}
The scalar mass matrix is, 
\beq
V_\mathrm{mass} = \frac{1}{2} \;U M^2 U^T,
\eeq
where 
\beq
U=
\left(
\begin{array}{cc}
h & S 
\end{array}\right),
\eeq
\bea
M^2\equiv\left(
\begin{array}{cc}
M^2_{11} & M^2_{12} \\
M^2_{12} & M^2_{22}
\end{array}\right)
=\left(
\begin{array}{cc}
3 \lambda v^2-\mu^2+x(a_1+a_2x)/2 & a_1 v/2+a_2 v x \\
a_1 v/2+a_2 v x & b_2+a_2 v^2/2+x(2b_3+3b_4 x)
\end{array}\right).
\label{massmatrix}
\eea
The mass eigenstates are 
\beq
\left(\begin{array}{c}
h_1\\ h_2
\end{array}\right) = 
\left(\begin{array}{cc}
\cos\theta & \sin\theta\\
-\sin\theta & \cos\theta
\end{array}\right)\,
\left(\begin{array}{c}
h\\ S
\end{array}\right).\ \ \ 
\label{mixdef}
\eeq
The physical  masses of $h_1$ and $h_2$ are $m_1^2$ and $m^2_2$, respectively:
\beq
m^2_{1,2}={1 \over 2}\left(M^2_{11}+M^2_{22}\mp \sqrt{(M^2_{11}-M^2_{22})^2+4 M^4_{12}}\right).
\label{m12}
\eeq
Note that the range of the mixing angle is $-\pi/4<\theta<\pi/4.$  We take $h_1$ to be the SM-like Higgs boson
with $m_{1}=126~GeV$.

As mentioned earlier, we are interested in the scenario where $(v,x)=(v_{EW},0)$ is the global minimum of the potential.  Hence, we require that the correct masses and mixing of the Higgs bosons are reproduced at this minimum:
\begin{eqnarray}
\left.\det M^2\right|_{\begin{subarray}{l} v=v_{EW}\\x=0\end{subarray}} = m_1^2m_2^2,
\quad \left.\Tr M^2\right|_{\begin{subarray}{l} v=v_{EW}\\x=0\end{subarray}}=m_1^2+m_2^2,
\quad {\rm and}\quad \left.\frac{2 M^2_{12}}{m_1^2-m_2^2}\right|_{\begin{subarray}{l} v=v_{EW}\\x=0\end{subarray}}=\sin2\theta.
\end{eqnarray}

From inspection, using  Eq.~\ref{mu2b1} and $x=0$, the mass matrix only depends on three combinations of parameters.  These can be solved for:\footnote{There are two solutions.  We choose this solution by using the further constraint that
$\lambda$ obtains the SM value, $\lambda=m_1^2/2v_{EW}^2$, in the limit $\theta\rightarrow 0$.}
\begin{eqnarray}
a_1&=&\frac{m_1^2-m_2^2}{v_{EW}}\sin2\theta,\nonumber\\
b_2+\frac{a_2}{2}v_{EW}^2&=&m_1^2\sin^2\theta+m_2^2\cos^2\theta,\nonumber\\
\lambda&=&\frac{m_1^2\cos^2\theta+m^2_2\sin^2\theta}{2v^2_{EW}}.
\label{paramdefs}
\end{eqnarray}
Our free parameters are then:
\begin{eqnarray}
m_1=126~{\rm GeV},\;m_2,\;\theta,\;v_{EW}=246~{\rm GeV},\;x=0,\; a_2,\;b_3,\;b_4.
\end{eqnarray}
Note that once we choose the masses, mixing, and vevs, there is little choice in the free parameters.  That is, all parameters are fully determined except $a_2,b_2,b_3,$ and $b_4$, and there is a relation between $b_2$ and $a_2$.

Since the singlet Higgs does not couple to the SM fermions and vector bosons, the couplings of $h_1$ and $h_2$ are determined by those of the neutral
component, $h$, of the Higgs doublet. From Eq. \ref{mixdef}, one can see that the coupling of $h_1$ to the SM fermions and vector bosons, normalized to 
the SM values, is suppressed by a factor $\cos\theta$, while the coupling of $h_2$ is suppressed by  $-\sin\theta$.

The self-interactions of the Higgs bosons in the basis of mass eigenstates $h_1$ and $h_2$ are,

\bea
\label{vcoup}
V_{\rm self}&\supset& {\lambda_{111} \over 3!} h_1^3 + {\lambda_{211} \over 2!} h_2 h_1^2 + {\lambda_{221} \over 2!} h_2^2 h_1 +{\lambda_{222} \over 3!} h_2^3 + \\ \nonumber
 &&{\lambda_{1111} \over 4!} h_1^4 +{\lambda_{2111} \over 3!} h_2 h_1^3 + {\lambda_{2211} \over 4} h_2^2 h_1^2+ {\lambda_{2221} \over 3!} h_2^3 h_1 +{\lambda_{2222} \over 4!} h_2^4.
\eea
The cubic and quartic couplings are listed in Appendix A.

The partial width of $h_2\rightarrow h_1h_1$ is then
\begin{eqnarray}
\Gamma(h_2\rightarrow h_1 h_1)=\frac{\lambda^2_{211}}{32\pi m_2}\sqrt{1-\frac{4 m_1^2}{m_2^2}}.
\label{gam2def}
\end{eqnarray}
Since the coupling of $h_2$ to other SM particles is suppressed by $\sin\theta$ we can write the total width\footnote{We neglect the partial width $h_2\rightarrow h_1h_1h_1$ since this is additionally suppressed by  three body phase space.}
\begin{eqnarray}
\Gamma(h_2)=\sin^2\theta\, \Gamma^{\rm SM}|_{m_2} +\Gamma(h_2\rightarrow h_1h_1),
\end{eqnarray}
where $\Gamma^{\rm SM}|_{m_2}$ is the SM Higgs total width evaluated at mass $m_2$.  In future calculations we use the results in Ref.~\cite{Gunion:1989we} to calculate $\Gamma^{\rm SM}$.

\subsection{Vacuum Structure}

Vacuum stability requires that the scalar potential must be positive definite 
as $\phi_0$ and $S$ become large.  The behavior of the potential at large values of
the fields is governed by the quartic interactions,
\beq
4\lambda \phi_0^4 + 2 a_2 \phi_0^2 S^2 +b_4 S^4 > 0\, .
\label{stability}
\eeq
We know that $\lambda$ and $b_4$ must both be positive since 
the potential needs to be stable along the axes $S=0$ or $\phi_0=0$.  Also, for $a_2>0$ the potential is clearly stable.  For $a_2<0$, 
rewrite Eq.~\ref{stability} as,
\beq
\lambda(2 \phi_0^2 + \frac {a_2}{2\lambda} S^2)^2 + (b_4- \frac{a_2^2}{4\lambda} ) S^4 > 0\, .
\label{stability2}
\eeq
Since the first term is positive definite, we obtain the stability bound
\begin{equation}
-2\sqrt{\lambda b_4}\le a_2.
\label{a2}
\end{equation}

Following the methods of Ref.~\cite{Espinosa:2011ax}, the extrema of Eq.~\ref{Vnew} for which $v\neq 0$ can be found:
\begin{eqnarray}
(v,x)=(v_{EW},0),\quad{\rm and}\quad (v,x)=(v_\pm,x_\pm)
\end{eqnarray}
where
\begin{eqnarray}
x_\pm&\equiv& \frac{v_{EW}(3a_1 a_2-8 b_3\lambda)\pm 8\sqrt{\Delta}}{4v_{EW}(4b_4\lambda-a_2^2)}\nonumber\\
v^2_\pm&\equiv& v_{EW}^2-\frac{1}{2\lambda}\left(a_1x_\pm + a_2 x^2_\pm\right),\nonumber\\
\Delta &=& \frac{v^2_{EW}}{64}\left(8b_3\lambda-3a_1a_2\right)^2-\frac{m_1^2m_2^2}{2}\left(4b_4\lambda-a_2^2\right)
\end{eqnarray}
For three real solutions to exist, we need $\Delta>0$ and $v_{\pm}^2>0$.  
There are also solutions for $v= 0$, which we include in the appendix.

\begin{figure}[tb]
\centering
\includegraphics[width=0.75\textwidth,clip]{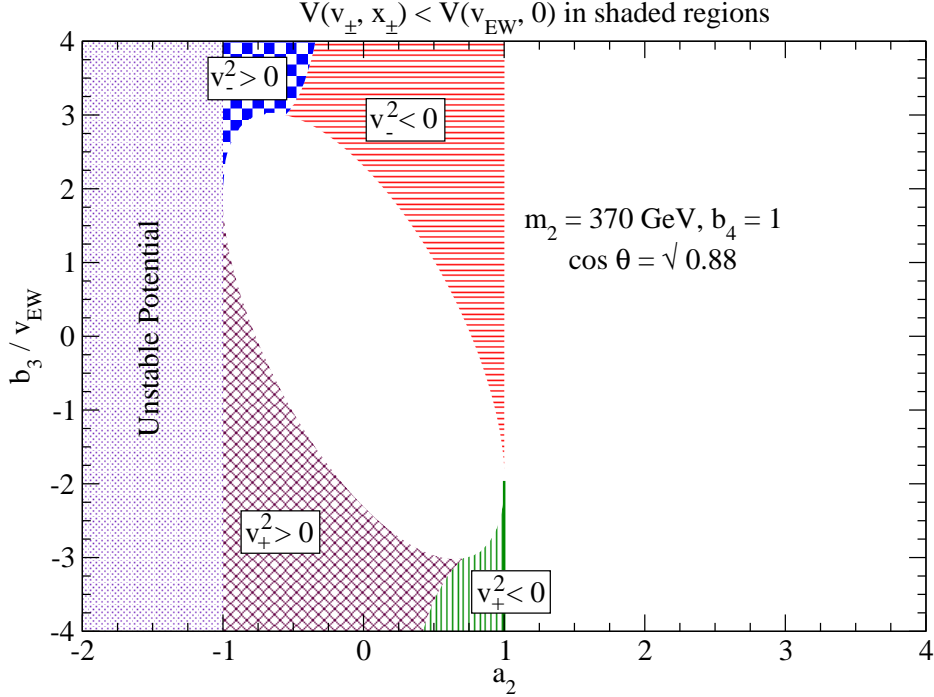}
\caption{Structure of the $v^2\neq 0$ vacua in the $b_3$ vs. $a_2$ plane for $m_{2}=370$~GeV, $b_4=1$, and $\cos\theta=\sqrt{0.88}$.  The different regions are where the $(v,x)=(v_{EW},0)$ minimum is the lowest lying (white region), $(v_-,x_-)$ is the lowest lying minimum with $v^2_-<0$ (red horizontal lines) and $v^2_->0$ (blue squares), and $(v_+,x_+)$ is the lowest lying minimum with $v^2_+<0$ (green vertical lines), and $v^2_+>0$ (maroon hatched region).}
\label{NonZero.fig}
\end{figure}

First, we analyze the $v^2\neq 0$ solutions.  For the global minimum to be $v=v_{EW}$ and $x=0$, the potential of Eq. \ref{Vnew} must satisfy
\begin{eqnarray}
V(v_{EW},0)<V(v_{\pm},x_{\pm}).
\end{eqnarray}
It can be shown that this occurs for,
\begin{eqnarray}
v_{EW}\mid8\lambda b_3-3 a_1 a_2\mid<6m_{1}m_{2}\sqrt{4b_4\lambda-a^2_2},\quad{\rm or}\quad 4b_4\lambda<a^2_2.~\label{vewmin.eq}
\end{eqnarray}
The vacuum structure of $v^2\neq 0$ is shown 
in Fig.~\ref{NonZero.fig} with $m_{2}=370$~GeV, $\cos\theta=\sqrt{0.88}$, and $b_4=1$.  
The region with $a_2 \lesssim -1$ does not satisfy the stability bound of Eq. \ref{a2}.
The white region is where the $(v,x)=(v_{EW},0)$  solution
is the lowest lying minimum with $v^2\neq 0$, as given in Eq.~\ref{vewmin.eq}.  The shaded areas show $b_3,a_2$ values where $V(v_-,x_-)<V(v_{EW},0)$ with $v_-^2<0$ (red horizontal lines) and $v_-^2>0$ (blue squares), and $V(v_+,x_+)<V(v_{EW},0)$ with $v_+^2<0$ (green vertical lines) and $v_+^2<0$ (maroon hatched lines).  All three solutions are never simultaneously minima. 
  
It can be shown that $(v_{EW},0)$ always corresponds to a minimum. Hence, this exhausts the possibilities for $v^2\neq 0$.  Since we require that the global minimum be real, we can also reject solutions for which $v^2_{\pm}<0$.  Hence, $v=v_{EW}$ and $x=0$ is the lowest lying real minimum with $v^2\neq0$ in the red-lined, green-lined, and white regions.  However, we must consider also the case $v=0$, which is discussed in the appendix. 

\begin{figure}[tb]
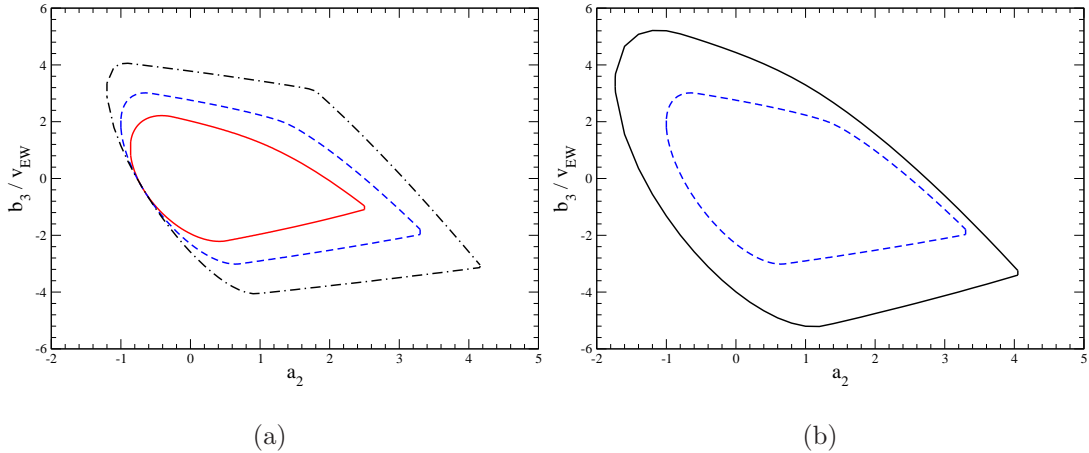

\begin{center}
\subfigure[]{
      \includegraphics[scale=0.3,clip]{EWmin.eps}\label{EWmin.a}}
\subfigure[]{\includegraphics[scale=0.3,clip]{EWmin_M370_b4.eps}\label{EWmin.b}}
\end{center}
\caption{ Constraints on the $(b_3$, $a_2)$ parameter space obtained
 by requiring that the global minimum is at 
$(v,x)=(v_{EW}=246~GeV,0)$. Regions enclosed by the lines are allowed.
Fig.~\ref{fig:EWmin}(a) shows the allowed regions with various values of $m_{2}$ for $b_4=1$. 
The solid (red), dashed (blue), and dash-dotted (black) represent $m_{2}=$ 270, 370, and 500 GeV, respectively.
Fig.~\ref{fig:EWmin}(b) shows the allowed regions with $b_4=1$ (blue dashed) and $b_4=3$ (black solid) for $m_{2}=370$ GeV.
The parameters used are $m_{1}=126$ GeV and $\cos\theta = 0.94$. }
\label{fig:EWmin}
\end{figure}

The final results for the allowed $(b_3,a_2)$ region with a global minimum at $(v,x)=(v_{EW},0)$ are shown in Fig.~\ref{fig:EWmin}.  This includes 
the analysis of the $v=0$ minima.  
Inside the contours $(v,x)=(v_{EW},0)$ is the global minimum. Fig.~\ref{EWmin.a} shows the dependence on the heavy scalar mass $m_2$, and Fig.~\ref{EWmin.b} shows the dependence on $b_4$.  Increasing $b_4$ and $m_{2}$  increases the upper bounds on $a_2$ slightly.  The difference in allowed regions between Figs.~\ref{NonZero.fig} and~\ref{EWmin.a} corresponds to the case where the 
$v=0$ minimum is the global minimum.

In Fig.~\ref{EWmin.a}, there is an interesting point on the contours that appears to be independent of $m_{2}$.  From Eq.~\ref{vewmin.eq}, this section of the contour arises from the inequality
\begin{eqnarray}
b_3^{min}\equiv\frac{3}{8\lambda v_{EW}}\left(a_1 a_2 v_{EW}-2 m_{1} m_{2} \sqrt{4b_4\lambda-a_2^2}\right)<b_3.
\end{eqnarray} 
The stationary points on this line can be found by solving $\partial b_3^{min}/\partial m_{2}=0$ for $a_2$.  Assuming $\sin\theta>0$, one of these solutions corresponds to
\begin{eqnarray}
a_2=-\sqrt{2 b_4}\cos\theta \frac{m_{1}}{v_{EW}},\quad{\rm and}\quad b_3=-\frac{3}{2}\sqrt{2b_4}\sin\theta m_{1},
\end{eqnarray}
which is independent of $m_{2}$.  This exactly corresponds to the degenerate point on the contours in Fig.~\ref{EWmin.a}.

It is clear from these results that both $a_2$ and $b_3$ are bounded for fixed masses, mixing, and $b_4$.  As we will see in Section \ref{sec:limits}, requiring perturbative unitarity bounds $b_4$.  Hence, all parameters are either determined by the masses and mixings of the Higgs sector or are bounded by theoretical considerations.  This will have a direct influence on the phenomenology of the singlet model  at the LHC.

%% file: di-Higgs.tex
\subsection{Results without a $Z_2$ Symmetry}
\begin{figure}[tb]
\begin{center}
\subfigure[]{\includegraphics[scale=0.5,clip]{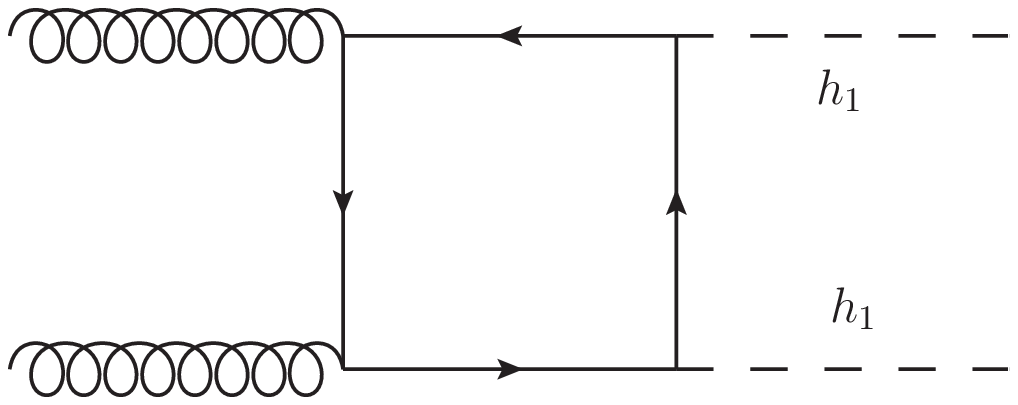}}
\subfigure[]{\includegraphics[scale=0.5,clip]{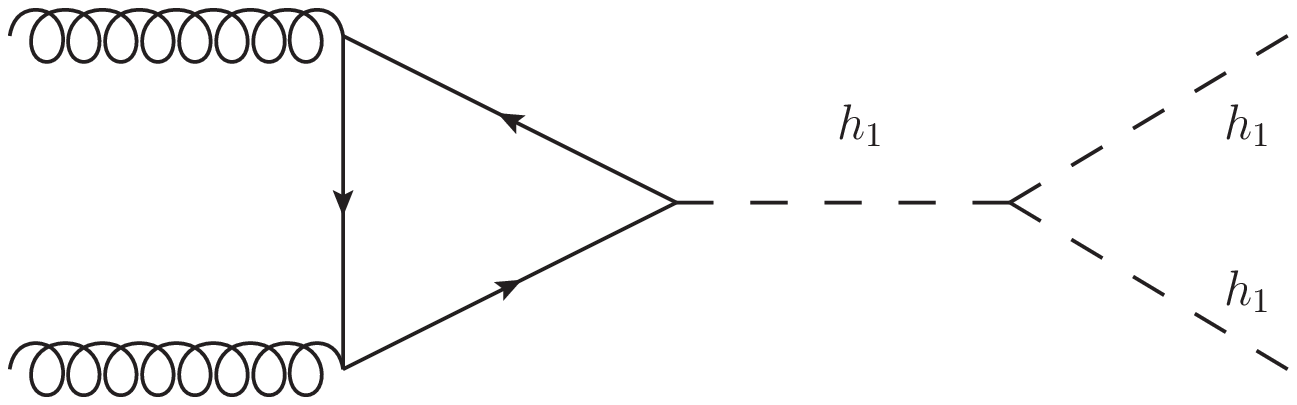}}
\subfigure[]{\includegraphics[scale=0.5,clip]{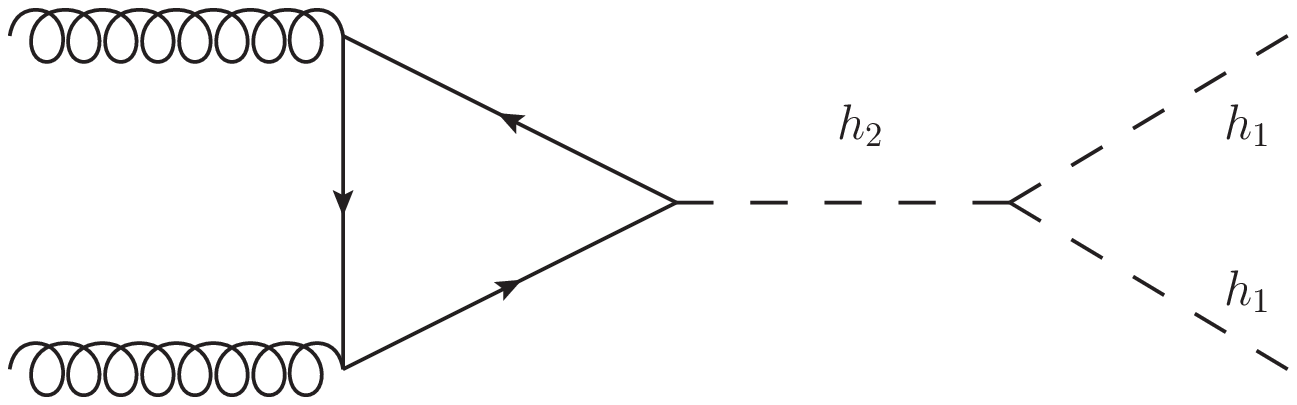}}
\end{center}
\caption{Representative diagrams for  di-Higgs production corresponding to (a) box diagram, (b) triangle diagram exchanging the light Higgs $h_1$, 
and (c) triangle diagram exchanging the heavy Higgs $h_2$. The solid lines stand for fermions, where top quark loops give the dominant contributions.}
\label{fig:gghh}
\end{figure}

\begin{figure}[tb]
\begin{center}
\includegraphics[scale=0.5,clip]{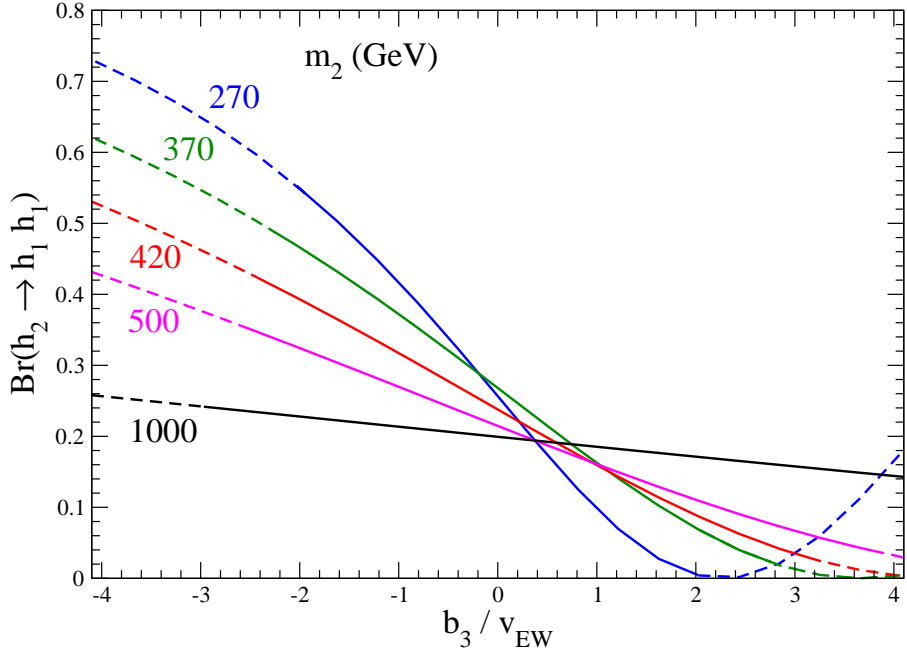}
\end{center}
\caption{ The branching ratio of $h_2 \to h_1 h_1$ as a function of $b_3$. 
The parameters used are $m_{1}=126$ GeV,  $\cos\theta = 0.94, a_2=0, v_{EW} =246$ GeV, and $b_4=1$. Lines from top to bottom
are $m_{2}=270, 370, 420, 500$, and 1000 GeV.
The solid (dashed) lines stand for regions that are allowed (excluded) by the requirement of EW stability. }
\label{fig:sigb3}
\end{figure}
\begin{figure}[tb]
\begin{center}
\subfigure[]{\includegraphics[scale=0.3,clip]{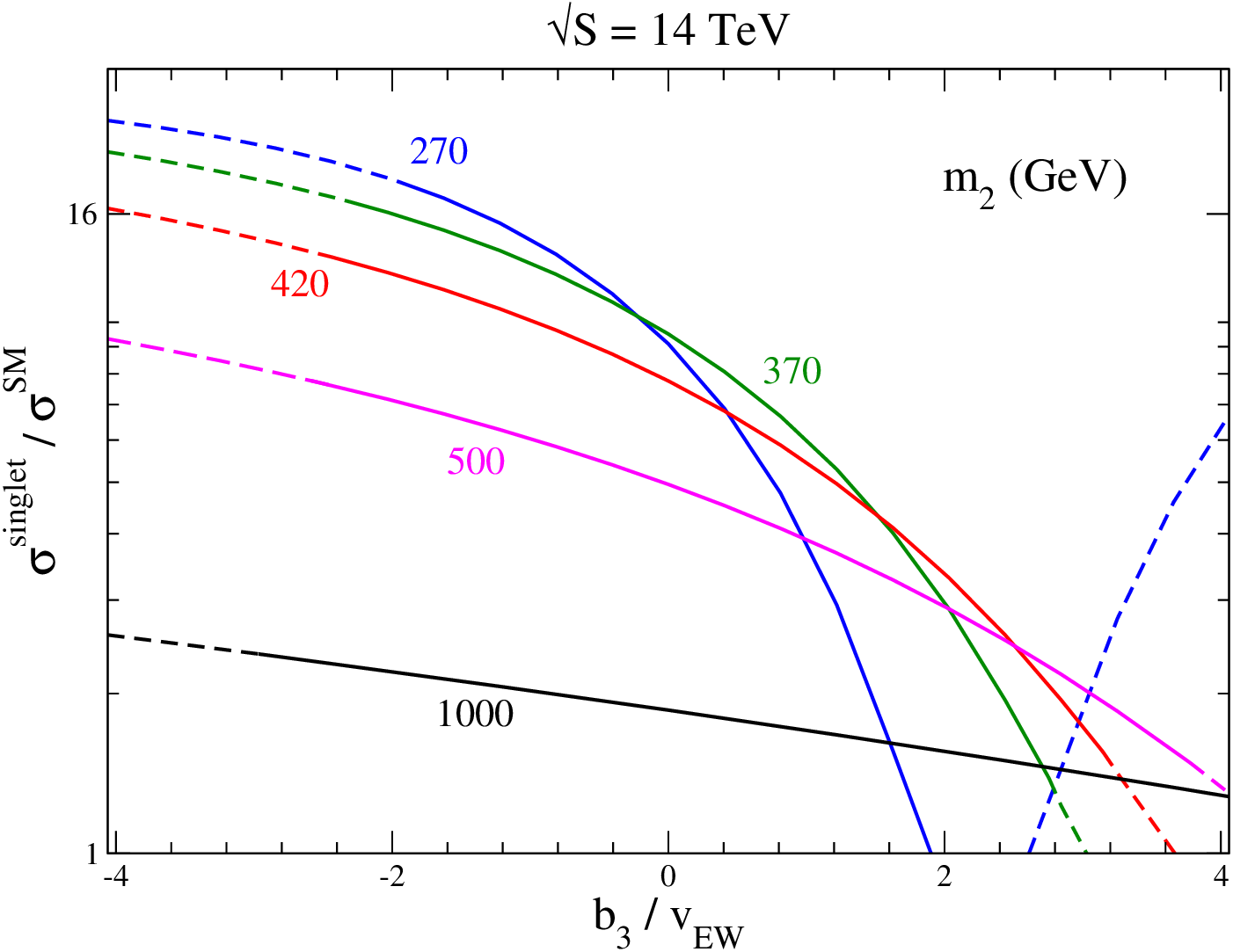}}
\subfigure[]{\includegraphics[scale=0.3,clip]{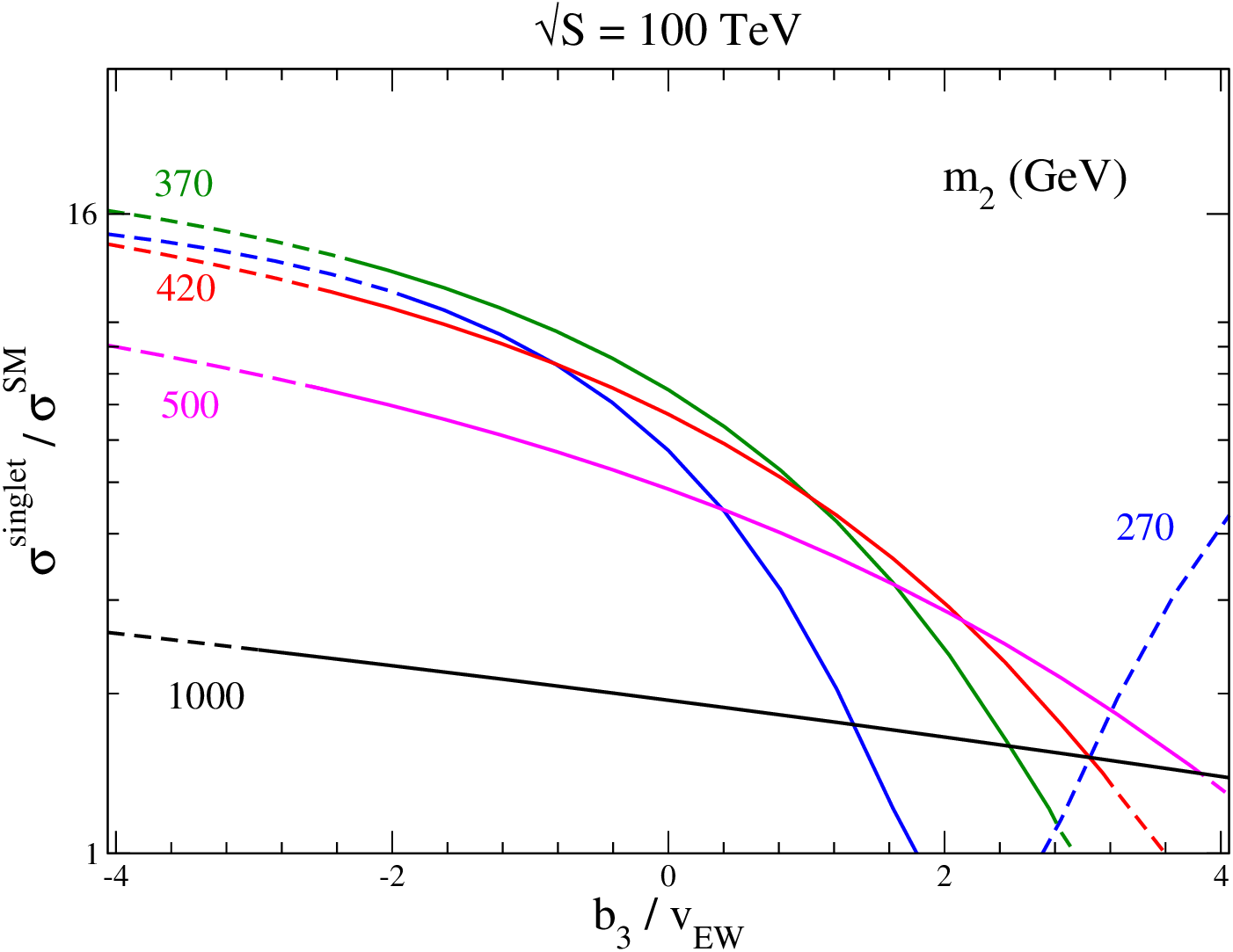}}
\end{center}
\caption{ The ratio of the di-Higgs cross section in the singlet model to that in the SM at (a) 
$\sqrt{S}=14~ TeV$ and (b) $\sqrt{S}=100~ TeV$ as a function of $b_3$. 
The parameters used are $m_1=126~ GeV$,  
$\cos\theta = 0.94$ , $a_2=0$, $v_{EW}=246~ GeV$, and $b_4=1$.
The solid (dashed) lines stand for 
regions that are allowed (excluded) by the requirement of EW stability. }
\label{fig:sighh3}
\end{figure}
\begin{figure}[tb]
\begin{center}
\subfigure[]{\includegraphics[scale=0.3,clip]{plot_sig_br_14tev.eps}}
\subfigure[]{\includegraphics[scale=0.3,clip]{plot_sig_br_100tev.eps}}
\end{center}
\caption{ The ratio of the di-Higgs cross section in the singlet model to that in the SM at (a) 
$\sqrt{S}=14~ TeV$ and (b) $\sqrt{S}=100 ~TeV$  as a function of 
the branching ratio of $h_2 \to h_1 h_1$. 
The parameters used are $m_{1}=126~GeV$,  $\cos\theta = 0.94$, 
$a_2=0$, $v_{EW}=246~ GeV$, and $b_4=1$.
The solid (dashed) lines stand for 
regions that are allowed (excluded) by the requirement of EW stability. $m_{2} = 270$ (brown), 
$420$ (red), and $1000~ GeV$ (black), respectively.}
\label{fig:sig1100}
\end{figure}
\begin{figure}[tb]
\begin{center}
\includegraphics[scale=0.5,clip]{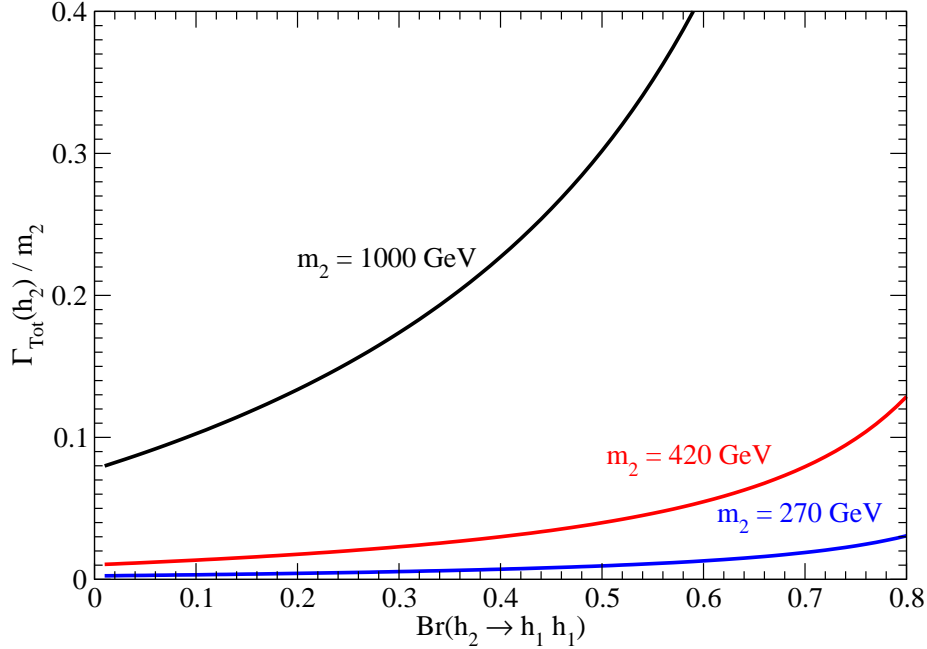}
\end{center}
\caption{The total width of $h_2$ as a ratio with $m_2$ vs. the branching ratio $h_2\rightarrow h_1h_1$.  The parameters used are $m_{1}=126~GeV$,  $\cos\theta = 0.94$, 
$a_2=0$, $v_{EW}=246~ GeV$, and $b_4=1$.  The masses are $m_2=270$~GeV (blue), $420$~GeV (red), and $1000$~GeV (black).}
\label{fig:WidthvsBR}
\end{figure}
We turn now to the results for di-Higgs production obtained by imposing the parameter restrictions described above to find 
the maximum 
enhancement possible in the $gg\rightarrow h_1 h_1$ channel relative to the SM rate.  Di-Higgs production proceeds through the diagrams shown in Fig.  \ref{fig:gghh}.   For $m_2\gtrsim 2 m_1$, it is possible to have a large
resonant enhancement from the diagram of Fig. \ref{fig:gghh}(c).   Our numerical results use CT12NLO PDFs with
$\mu=M_{h_1h_1}$.  We normalize many of our plots to the LO SM predictions, $\sigma(gg\rightarrow 
h_1 h_1)\mid_{SM}=
15~fb~ (0.6~pb)$ at $\sqrt{S}=14~TeV (100~TeV$).\footnote{Radiative corrections in the SM are large, typically a 
factor of $\sim 2$ enhancement\cite{Dawson:1998py,Grigo:2014jma,deFlorian:2014rta},
and are not included here since they are simply an overall normalization
factor to the results we present.} 

From the mass matrix in Eq. \ref{massmatrix},  we know that varying $b_3$ does not change $m_{1}$, $m_2$ and the mixing angle $\theta$. 
In contrast, one can observe that $\lambda_{211}$ in Eq.~\ref{self} is a function of $b_3$.
In Fig. \ref{fig:sigb3}, we show the dependence  on $b_3$ of  the branching ratio of the heavier Higgs, $h_2$, into the SM-like Higgs, $h_1$.  For $b_3$ small, the branching ratio has little dependence on $m_2$, while for large
$b_3$, the branching ratio can be large and depends significantly on $b_3$.     The dotted curves  represent regions where the parameters do not correspond to a global electroweak minimum.  We see then that for a given mass this constraint corresponds to an upper limit on the branching ratio ${\rm Br}(h_2\rightarrow h_1h_1)$.

To understand the features of Fig. \ref{fig:sigb3},
 use the solutions in Eq.~\ref{paramdefs} to rewrite
\begin{eqnarray}
\lambda_{211}=\sin\theta\left[-\frac{2 m_1^2+m_2^2}{v_{EW}}\cos^2\theta-a_2 v_{EW}\left(1-3\cos^2\theta\right)+b_3 \sin(2\theta)\right].
\label{lam211phys}
\end{eqnarray} 
From this we see that $b_3\sin(2\theta)$ and $m_2$ make opposite sign contributions to $\lambda_{211}$.  Hence, for $b_3\sin(2\theta)<0$, they constructively contribute to $\lambda_{211}$.  The major feature of this region in Fig.~\ref{fig:sigb3} is then understood by noting that the partial widths of $h_2$ into $h_1$, $W$s, and $Z$s scale like
\begin{eqnarray}
\Gamma(h_2\rightarrow h_1h_1)\propto\sin^2\theta\, m_2,\quad{\rm and}\quad \Gamma(h_2\rightarrow W^+W^-/ZZ)\propto \sin^2\theta\,m_2^3.
\end{eqnarray}
 Hence, as the mass of $h_2$ increases the partial widths into $W$s and $Z$s grow much 
 more quickly than the partial width into $h_1h_1$.  The branching ratio ${\rm Br}(h_2\rightarrow h_1h_1)$ therefore decreases with mass.

The region for $b_3\sin(2\theta)>0$ is slightly more involved. Using Eq.~\ref{lam211phys}, the triple coupling $\lambda_{211}$ goes to zero when
\begin{eqnarray}
b_3\sin(2\theta)=\frac{2m_1^2+m_2^2}{v_{EW}}\cos^2\theta + a_2 v_{EW}\left(1-3\cos^2\theta\right).
\end{eqnarray}
We see that for smaller $m_2$ the zero corresponds to smaller $b_3\sin(2\theta)$.  As $b_3\sin(2\theta)$ goes from negative to positive, the smaller $m_2$ values turn over and approach zero more quickly than the larger $m_2$.  This is the behaviour we see in Fig.~\ref{fig:sigb3}.  Note that for our representative parameters, we have $\theta>0$, so the sign of $b_3\sin(2\theta)$ is the same as $b_3$.

In Fig. \ref{fig:sighh3}, we plot 
 the dependence of 
the ratio of the di-Higgs production cross section in the singlet model to that in the SM. 
 In this type of model, the double Higgs production cross section can reach up to $\mathcal{O}(10)$ times that of 
the SM with $58\%\gtrsim {\rm Br}(h_2 \to h_1 h_1) \gtrsim 28$\%.  Interestingly, the enhancement does not grow as $\sqrt{S}$ is
increased from $14~TeV$ to $100~TeV$, although of course the total rate is increased.  Both the SM and singlet rates are dominated by gluon fusion production; hence, both rates are similarly increased between $14$ and $100$~TeV.  
\begin{figure}[tb]
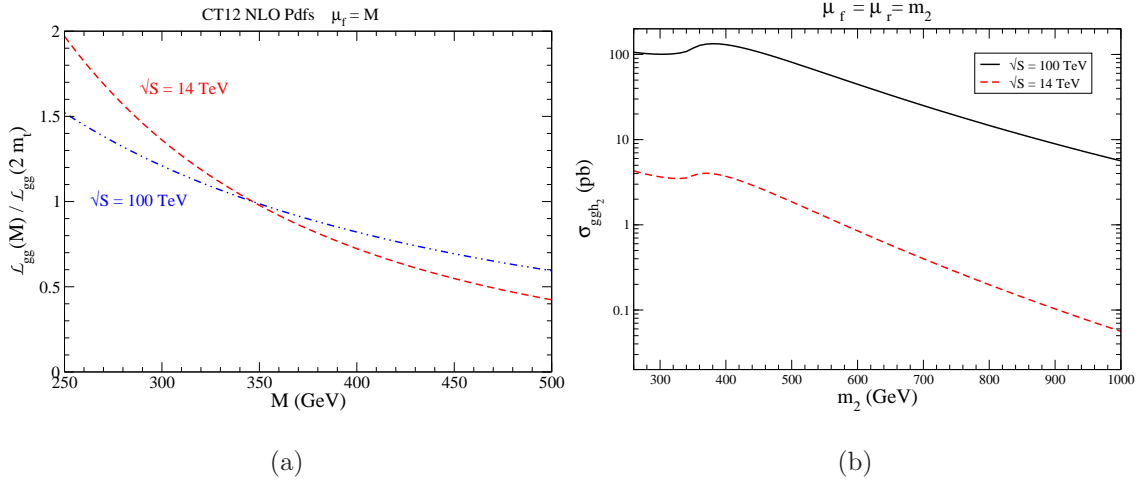

\begin{center}
\subfigure[]{\includegraphics[scale=0.3,clip]{ggPartLumNorm.eps}}
\subfigure[]{\includegraphics[scale=0.3,clip]{ggh_sig_m2_mu=m2.eps}}
\end{center}
\caption{ (a) Gluon gluon luminosity at $\sqrt{S}=14$ and $100~TeV$ as a function of invariant mass, $M$.
(b) Resonant contribution from $gg\rightarrow h_2$, evaluated at a scale, $\mu=m_2$ with $\cos\theta=.94$.}
\label{fig:gglum}
\end{figure}

The di-Higgs enhancement depends on the production cross section of $h_2$ and the branching ratio of $h_2\rightarrow h_1h_1$.  Since the production cross section of lower mass states is generically larger than that of high mass states, $m_2=270$~GeV has the largest enhancement for $b_3<0$.  For $b_3>0$, it is possible for the branching ratio of $h_2\rightarrow h_1 h_1$ to go to zero.  The behaviour of the enhancement in this region closely follows the discussion of Fig.~\ref{fig:sigb3}.
  For $\sqrt{S}=100~TeV$ and $b_3 < 0$ (Fig. 5(b)), the cross section for $m_2=270~GeV$ drops below that of $m_2=370~GeV$.  As to be discussed later, this is due to specific properties of di-Higgs production.

In Fig.~\ref{fig:sig1100} we show the the enhanced di-Higgs ratio as a function of the $h_2\rightarrow h_1h_1$ branching ratio.  If the narrow width approximation holds and the production cross section $h_2$ is sufficiently larger than the SM di-Higgs rate, we have
\begin{eqnarray}
\sigma(pp\rightarrow h_1 h_1)\approx \sigma(pp\rightarrow h_2){\rm Br}(h_2\rightarrow h_1 h_1).
\end{eqnarray}
Hence, we would expect this dependence to be a straight line, as seen for $m_2=270$ and $420$ GeV.  However, we see that this is not the case for $m_2=1000$~GeV.  In Fig.~\ref{fig:WidthvsBR} we show the ratio of the total width of $h_2$ and $m_2$ as a function of the branching ratio of $h_2\rightarrow h_1 h_1$.  As can be seen for $m_2=1000$~GeV, the width is always large and the narrow width approximation is poor.  This explains why the $m_2=1000$~GeV line in Fig.~\ref{fig:sig1100} is not straight.  Also, as the branching ratio of $h_2\rightarrow h_1h_1$ increases, the total width become larger.  This is due to the partial width $h_2\rightarrow h_1h_1$ becoming large, since the partial widths into $W$ and $Z$ boson is fixed by the mass $m_2$ and mixing angle $\theta$.

\begin{figure}[t]
\begin{center}
\includegraphics[scale=0.5,clip]{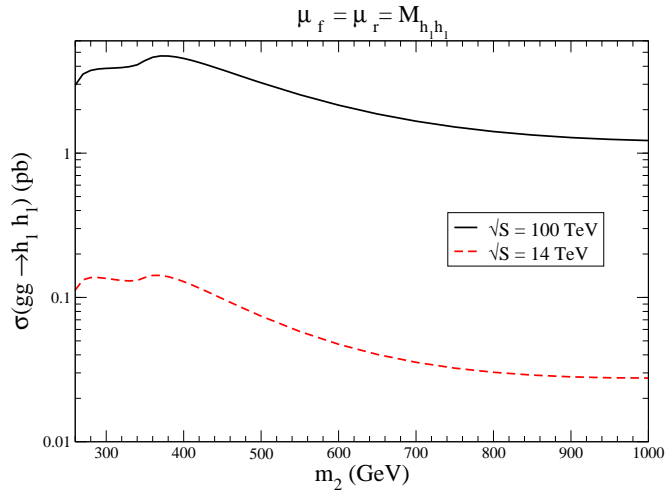}
\end{center}
\caption{Total cross section for $gg\rightarrow h_1h_1$ as a function of $m_2$ for $b_3=a_2=0$, $b_4=1$, 
and $\cos\theta=.94$.}
\label{m2def}
\end{figure}

In Fig. \ref{fig:sig1100}, it is interesting to note that the enhancement for $m_2=420~GeV$ is larger than that
for $270~GeV$ at $\sqrt{S}=100~TeV$.   This can be understood from the parton luminosity 
plot of Fig. \ref{fig:gglum}(a), where
we show the gluon-gluon parton luminosity (normalized to that at $2m_t$).  The $\sqrt{S}=14~TeV$ luminosity
falls much more quickly as a function of  invariant mass  than does the 
corresponding luminosity at $\sqrt{S}=100~TeV$.   We compare this
with the resonant production of $gg\rightarrow h_2$ in Fig. \ref{fig:gglum}(b)
 and observe  that at $\sqrt{S}=100~TeV$ the resonant
enhancement at  the $t {\overline t}$ threshold is more important than at $\sqrt{S}=14~TeV$.  Finally, we show
the dependence on $m_2$ of the full cross section for $gg\rightarrow h_1h_1$  in Fig. \ref{m2def}.  The resonant structure near $2m_t$ is clearly visible. 
\vskip .5in
\subsection{The $Z_2$ Limit}
It may be necessary in certain models 
 to impose a $Z_2$ symmetry on the potential under  which $S$ is odd and  $H$ is even.  This may be motivated from a dark matter perspective, where $S$ is a dark matter particle, or the point of view of a complex hidden sector.  The potential for this case can be obtained in the limit $a_1,b_1,b_3\rightarrow 0$.  If the $Z_2$ remains unbroken, there is no resonance enhancement in di-Higgs production, since the $S\rightarrow hh$ decay breaks the $Z_2$
 symmetry and there is no mixing between $S$ and $h$. We ignore this case. 
 However, the $Z_2$ symmetry  may be broken by a vev of $S$.  
 Unlike the case outlined above, the vev of $S$ is then physically meaningful and we cannot set $\vev{S}=x=0$ arbitrarily.    The $Z_2$ symmetric potential is,
 \begin{equation}
 V(H,S)=-\mu^2 H^\dagger H
 +\lambda(H^\dagger H)^2+{a_2\over 2}H^\dagger H S^2 +{b_2\over 2}S^2 +{b_4\over 4}S^4\, .
 \end{equation}
 We shift the fields in the usual manner to find the $h_2 h_1 h_1 $ coupling in the $Z_2$ symmetric limit\cite{Pruna:2013bma},
 \begin{equation}
 \lambda_{211}^{Z_2}=a_2\biggl[vs(2c^2-s^2)-xc(2s^2-c^2)\biggr]-
 6\lambda v c^2s+6b_4 x c s^2\, .
 \label{zl122}
 \end{equation}
 In the limit $x=0$ and $a_1,b_1$, and $ b_3=0$, Eq. \ref{zl122} is in agreement with Eq. \ref{self}.
 We impose the conditions of positivity of the potential, $\lambda>0, b_4>0$ and $4 \lambda b_4-a_2^2 > 0$
 (Eq. \ref{stability}) and require the couplings to be perturbative, $a_2,b_4,\lambda< 4 \pi$.
 
\begin{figure}[tb]
\begin{center}
{\includegraphics[scale=0.6,clip]{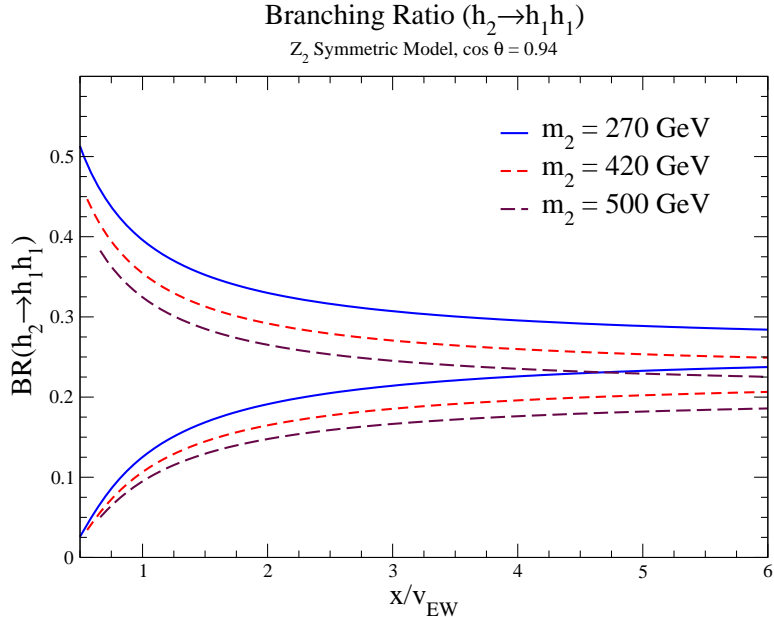}}
\end{center}
\caption{ The branching ratio of $h_2\rightarrow h_1 h_1$ in a $Z_2$ symmetric
model as a function of the vev of the singlet field, $x$. The upper (lower) branches of the
curves correspond to negative (positive) values of $\sin\theta$.   }
\label{fig:z2brs}
\end{figure}

 The physical parameters are taken as,
 \begin{equation}
 m_1,m_2,\cos \theta\equiv c, v_{EW}, x\, .
 \end{equation}
  Using Eqs. \ref{zl122}  and \ref{gam2def},  the branching ratio for $h_2\rightarrow h_1 h_1$ can be found and is 
 shown in Fig. \ref{fig:z2brs}.
 Comparing with Fig. \ref{fig:sig1100},
  it is apparent that the branching ratios are similar in the models with and without
 the $Z_2$ symmetry for large values of $x/v_{EW}$, where
 the
 branching ratio asymptotes
 to around $BR(h_2\rightarrow h_1 h_1)\sim 0.3$. The branching ratio $h_2\rightarrow h_1 h_1$ 
 appears to have little
 discriminating power between the $Z_2$ symmetric and non-symmetric
 potentials.   

%% file: limits.tex
There are a number of well known experimental and theoretical limits on the Higgs singlet model, which we briefly review in this section.

\subsection{Experimental Limits}

From the direct measurements of the Higgs coupling strengths, 
 ATLAS  \cite{atlasNP} places a constraint on the mixing angle, $\theta$,
 of the singlet model, where $\cos^2\theta \le 0.88$ has been excluded at 95\% CL.  This limit assumes that there is no branching ratio to
 invisible particles.
Here we take the upper  limit of $\sin^2 \theta\le 0.12$ as a representative point.
Direct searches for the heavy Higgs ($h_2$) decaying into $W^+W^-$ and $Z Z$ from ATLAS and CMS \cite{ATLAS-CONF-2013-067,Chatrchyan:2013yoa} can also give bounds on $\sin^2\theta$ 
with $\sin^2\theta \lesssim 0.2$ for $m_2\sim 200-400$ GeV and $\sin^2\theta \lesssim 0.4$ for $m_2 \sim 600$ GeV. However, these constraints are not 
as strong as the ATLAS limit from the Higgs coupling strengths.

\begin{figure}[t]
\begin{center}
{\includegraphics[scale=0.6,clip]{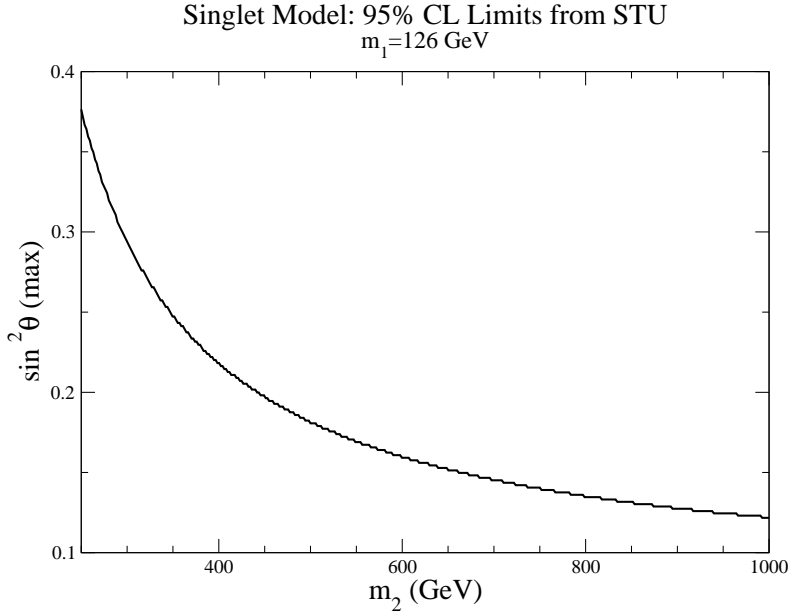}}
\end{center}
\caption{ Constraints on the mixing angle, $\sin\theta$, as a function of the mass of the heavier
Higgs scalar, $m_2$, from fits to the oblique parameters, $S$ and $T$.}
\label{fig:singlet}
\end{figure}

The existence of a Higgs singlet which mixes with the SM Higgs boson is also restricted by electroweak
precision observables.  A fit to the oblique parameters, $S$ and $T$ (fixing $U$ to be $0$), is
shown in Fig. \ref{fig:singlet} \cite{Dawson:2009yx,Profumo:2007wc}.  We see that limits from the oblique parameters are not competitive with the
ATLAS limit from the Higgs coupling strengths.

\begin{figure}[tb]
\begin{center}
{\includegraphics[scale=0.5,clip]{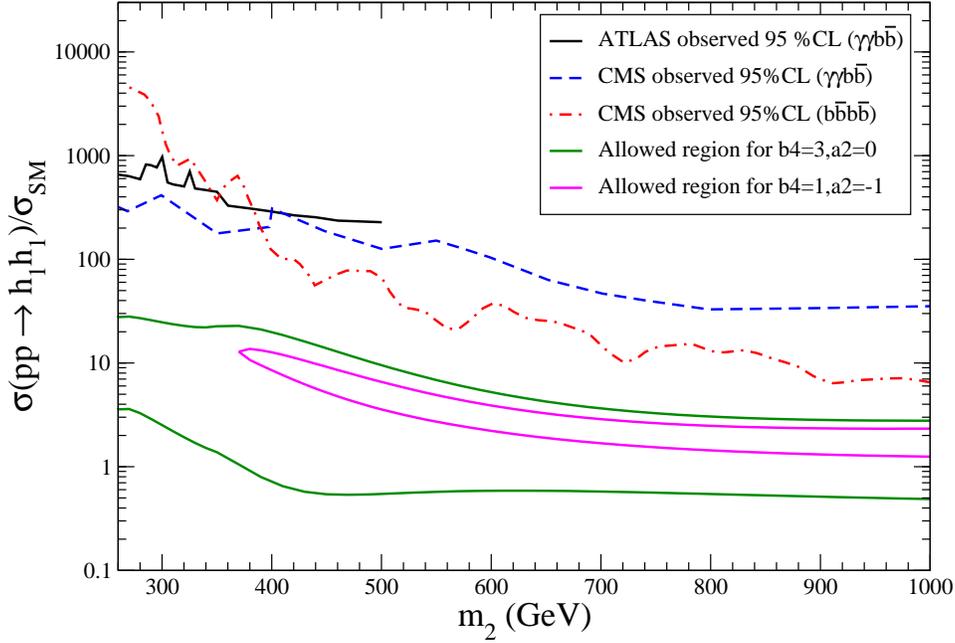}}
\end{center}
\caption{Observed $95\%$ CL upper limits at $\sqrt{S}=8$ TeV with an integrated luminosity of 20 fb$^{-1}$ on the
resonant di-Higgs production cross section from ATLAS in the $\gamma\gamma b {\bar b}$ channel (black solid), 
CMS in the $\gamma\gamma b {\bar b}$ channel (blue dashed) and CMS  in the $b {\bar b}b {\bar b}$  channel (red dot-dashed), normalized to the leading order cross section predicted by
the SM, and the regions allowed by the requirement that the electroweak minimum
be a global minimum  for $(b_4,a_2)=(3,0)$ (green solid) and $(b_4,a_2)=(1,-1)$ (magenta solid).}
\label{fig:08lhc}
\end{figure}

ATLAS and CMS have obtained upper bounds on the 
cross section for  the resonant production of SM  Higgs bosons pairs through
the process $pp\rightarrow h_2^*\rightarrow h_1 h_1$ 
in the $\gamma\gamma b {\bar b}$ \cite{Aad:2014yja,CMS-PAS-HIG-13-032}  and $b {\bar b}b {\bar b}$ \cite{CMS-PAS-HIG-14-013} channels
at a center-of-mass energy of $\sqrt{S}=8~ TeV$
 with an integrated luminosity of ~$20~ fb^{-1}$ as summarized in Fig.~\ref{fig:08lhc}.
In the low mass region the $\gamma\gamma b {\bar b}$ channel gives a stronger bound as opposed to a weaker bound obtained in the $b {\bar b}b {\bar b}$ 
channel due to 
the large QCD background. However, the limit from the $b {\bar b}b {\bar b}$ channel becomes more 
constraining above  $m_2 \sim 400$ GeV.

We compare the experimental upper limits on the production cross sections 
for resonant di-Higgs production with $m_2$ between $270~ GeV$ and $1~TeV$,
 normalized to the leading order cross section predicted by
the SM, with the range of allowed cross sections consistent with the requirement that
the parameters correspond to a global electroweak minimum.  (The allowed region is between
the curves).
 Two sets of parameter points $(b_4,a_2)=(3,0)$ and $(b_4,a_2)=(1,-1)$ are considered. The former has a larger value of $b_4$
and hence the bound is less stringent as 
illustrated in Fig.~\ref{fig:EWmin}(b). 
The lower limit of the allowed region on $m_2$, which starts at $ m_2\sim 370$ GeV, for $(b_4,a_2)=(1,-1)$ can be explained by Eq.~\ref{a2} as due to  the vacuum stability constraint. 
Plugging in $\lambda$ defined in Eq.~\ref{paramdefs},  one can obtain the lower limit for $m_2^2$ for a  given
 $b_4$ and negative $a_2$,
\begin{equation}
m_2^2 \ge \frac{1}{\sin^2\theta}\left(\frac{a^2_2}{2 b_4}v^2_{EW}-m_1^2 \cos^2\theta\right).
\end{equation}

\begin{figure}[tb]
\begin{center}
{\includegraphics[scale=0.5,clip]{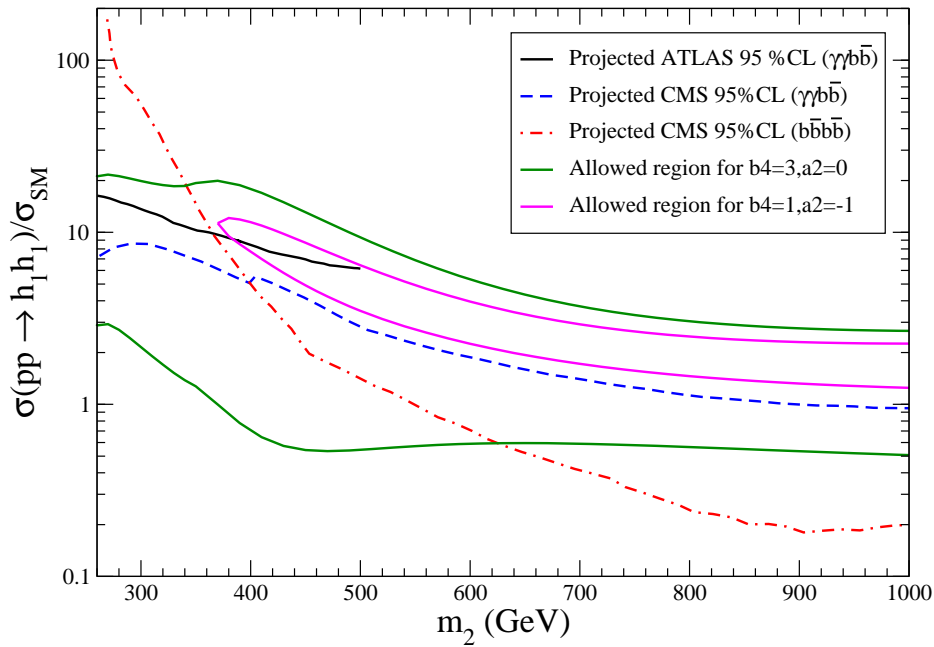}}
\end{center}
\caption{Projected $95\%$ CL upper limits at $\sqrt{S}=14$ TeV with an integrated luminosity of $300~ fb^{-1}$ on the production cross section from the
ATLAS $\gamma\gamma b {\bar b}$ channel (black solid), CMS $\gamma\gamma b {\bar b}$ (blue dashed) and CMS $b {\bar b}b {\bar b}$ (red dot-dashed), normalized to the leading order cross section predicted by
the SM, and the regions allowed by the requirement that the  electroweak minimum be a global minimum for $(b_4,a_2)=(3,0)$ (green solid) and $(b_4,a_2)=(1,-1)$ (magenta solid).}
\label{fig:14lhc}
\end{figure}

Throughout the $m_2 < 1~TeV$ mass range, the constraints derived from the global electroweak 
minimum requirement are always stronger than those currently available
 from the LHC experiments at $\sqrt{S}=8$ TeV. 
We make  naive projections for the expected 
constraints at 
the  LHC at $\sqrt{S}=14$ TeV with an integrated luminosity of 
$300~ fb^{-1}$ by rescaling the expected $95\%$ CL upper 
limits at $\sqrt{S}=8$ TeV with an integrated luminosity of $20~ fb^{-1}$,
using the ratios of gluon-gluon luminosities (evaluated at the scale  $2m_1$)  given in 
Ref. \cite{Quigg:2011zu}. 
As shown in Fig.~\ref{fig:14lhc}, the projected bounds from the CMS $\gamma\gamma b {\bar b}$ channel can rule out the entire parameter space 
where the electroweak minimum is a global minimum
for $(b_4,a_2)=(1,-1)$ and 
can exclude much
 of the allowed region for $(b_4,a_2)=(3,0)$. Moreover, the projected limits from the CMS $b {\bar b}b {\bar b}$ channel can potentially exclude  
 the entire  parameter space allowed by the electroweak  minimum
requirement
 for $(b_4,a_2)=(1,-1)$ and rule out two thirds of the allowed region  in the high mass range for $(b_4,a_2)=(3,0)$.

\subsection{Unitarity}
The coefficients of the potential cannot be too large or perturbative unitarity will be violated in the $h_ih_j$ scattering
processes \cite{Lee:1977eg}.  The simplest
limit comes from the high energy scattering of $h_2 h_2\rightarrow h_2 h_2$, where the $J=0$
partial wave is,
\begin{equation}
a_0(h_2h_2\rightarrow h_2 h_2)\rightarrow_{s>>m_2^2} {3b_4\over 8\pi}\, .
\end{equation}
Requiring $\mid a_0\mid <{1\over 2}$ yields $\mid b_4\mid \le 4.2$.  Limits from a coupled channel analysis of $h_ih_j$ scattering show that for small  $\sin\theta$, multi-TeV scale masses are allowed
for $m_2$ \cite{Pruna:2013bma}.

Similarly, we can consider the $h_1 h_1\rightarrow h_1 h_1$ scattering to find the $J=0$ partial wave.
\begin{equation}
a_0(h_1h_1\rightarrow h_1 h_1)\rightarrow_{s>>m_1^2} {3\lambda\over 8\pi}\, .
\end{equation}
Then using Eq.~\ref{paramdefs} and $\mid a_0 \mid <{1\over 2}$, an upper limit on $m_2$ can be found:
\begin{equation}
m^2_2 <\frac{1}{3\sin^2\theta}\left(8\pi v_{EW}^2-3 m_1^2 \cos^2\theta\right).
\end{equation}
For $\cos^2\theta = 0.88$ and $m_1=126$~GeV, this limit is $m_2\lesssim 2$~TeV.

%% file: conclusions.tex
We studied resonance enhancement of di-Higgs production in a generic singlet extended Standard Model.  By imposing conditions on the masses, mixing, and vacuum expectation values of the bosons we were able to identify the three parameters that are left free.  These three parameters were then bounded by unitarity constraints and the requirement that the electroweak symmetry breaking minimum be the global minimum.  With these constraints,  ${\rm Br}(h_2\rightarrow h_1 h_1)$ is bounded from above.  Hence, we found that theoretical considerations bound the di-Higgs production in this model and that the theoretical constraints are more stringent than the current limits from direct searches for $h_1h_1$.  We then provided predictions for the cross sections and branching ratios for $\sigma(pp\rightarrow h_2\rightarrow h_1 h_1)$ at both the 14 TeV LHC and a 100 TeV collider.  
The di-Higgs production enhancement can be as large as a factor of $\sim 18 (13)$ for $m_2=270 (420)$ GeV relative to the SM rate at 14 TeV
for parameters corresponding to a global EW minimum.



%% file: appendixa.tex
The cubic and quartic couplings in Eq.~\ref{vcoup} are listed below,

\bea
\lambda_{111}&=& 2 s^3 b_3+{3 a_1 \over 2}s c^2+3 a_2 s^2 c v + 6 c^3\ \lambda v,\nonumber \\
\lambda_{211}&=& 2 s^2cb_3+{a_1 \over 2}c(c^2-2s^2)+(2c^2-s^2)sva_2 -6\lambda sc^2v
\nonumber \\
\lambda_{221}&=& 2 c^2sb_3+{a_1 \over 2}s(s^2-2c^2)-(2s^2-c^2)cva_2 +6\lambda cs^2v
\nonumber \\
\lambda_{222}&=& 2 c^3 b_3+{3 a_1 \over 2}c s^2-3 a_2 c^2 s v - 6 s^3\ \lambda v,\nonumber \\
\lambda_{1111}&=&6(\lambda c^4+a_2s^2c^2+b_4 s^4)
\nonumber \\
\lambda_{2111}&=&6sc(b_4 s^2+ {a_2 \over 2}(1-2s^2) - \lambda c^2)
\nonumber \\
\lambda_{2211}&=& 6s^2c^2(-a_2+b_4+\lambda)+a_2
\nonumber \\
\lambda_{2221}&=&6sc(b_4 c^2+ {a_2 \over 2}(1-2c^2) - \lambda s^2)
\nonumber \\
\lambda_{2222}&=&  6(s^2c^2 a_2+c^4 b_4 +\lambda s^4)
\label{self}
\, ,
\end{eqnarray}
and we abbreviate $s=\sin\theta$, $c=\cos\theta$.  We assume $\sin\theta>0$.  Flipping
the sign of $\sin\theta$ is equivalent to reversing the sign of $b_3$, as is apparent
in Eq. \ref{self}.  Note that several couplings are related by a transformation 
$c\to -s$ and $s\to c$. To understand this, one can see that Eq.~\ref{mixdef} is invariant under $c \to -s, s \to c, h_1 \to h_2$, and $h_2 \to -h_1$.
This implies Eq.~\ref{vcoup} is also invariant under such transformations. As a result, the couplings
$\lambda_{111}, \lambda_{221}, \lambda_{1111},$ and $\lambda_{2222}$ are transformed into 
$\lambda_{222}, \lambda_{211}, \lambda_{2222},$ and $\lambda_{1111}$, respectively  after the replacement
$c\to -s$ and $s \to c$ while  $\lambda_{2211}$ remains invariant.
Similarly, $\lambda_{211}, \lambda_{222}, \lambda_{2111},$ and $ \lambda_{2221}$ are transformed into
$\lambda_{221}, \lambda_{111}, \lambda_{2221},$ and $\lambda_{2111}$, respectively under  $c\to -s$ and $s \to c$
up to a minus sign because they are associated with odd numbers of $h_2$.
In the small angle limit, to ${\cal O}(s^2)$,
\bea
\lambda_{111}&\rightarrow & 6\lambda v+{3\over 2} a_1s+ 3vs^2(a_2-3\lambda)
\nonumber \\
\lambda_{211}&\rightarrow & {a_1\over 2} +s v(-6\lambda +2a_2)+{s^2\over 4}(8b_3-7a_1)
\nonumber \\
\lambda_{221}&\rightarrow& 2 sb_3-a_1 s+(1-\frac{7}{2}s^2)va_2 +6\lambda s^2v
\nonumber \\
\lambda_{222}&\rightarrow& (2-3s^2) b_3+{3 a_1 \over 2} s^2-3 a_2 s v,
\nonumber \\
\lambda_{1111}&\rightarrow &6 \lambda-6s^2(2\lambda-a_2)
\nonumber  \\
\lambda_{2111}&\rightarrow&3s( a_2 - 2\lambda)
\nonumber \\
\lambda_{2211}&\rightarrow & a_2 +6s^2(-a_2+b_4+\lambda)
\nonumber \\
\lambda_{2221}&\rightarrow&3s(2b_4- a_2 )
\nonumber \\
\lambda_{2222}&\rightarrow&  6b_4+6s^2(a_2-2b_4)\, .
\eea

%% file: appendixb.tex
We now evaluate the extrema of the potential with $v=0$.  These are found by evaluating the extrema of Eq.~\ref{Vnew}.  The solutions for $\langle S\rangle$ are,
\begin{eqnarray}
x^0_1&=& \frac{(2\,b_3-\kappa^{1/3})^2-12\,b_2 b_4}{6\,b_4 \kappa^{1/3}}+\frac{b_3}{3\,b_4}\nonumber\\
x^0_2&=& \frac{(2\,b_3-e^{2 i \pi/3}\kappa^{1/3})^2-12\,b_2 b_4}{6\,b_4 e^{2 i \pi/3} \kappa^{1/3}}+\frac{b_3}{3\,b_4}\nonumber\\
x^0_3&=& \frac{(2\,b_3-e^{4 i \pi/3}\kappa^{1/3})^2-12\,b_2 b_4}{6\,b_4 e^{4 i \pi/3} \kappa^{1/3}}+\frac{b_3}{3\,b_4},
\end{eqnarray}
where we have defined,
\begin{eqnarray}
\kappa &=& -4\,b_3(2\,b_3^2-9\,b_2 b_4) + 27\,a_1 b_4^2 v_{EW}^2 +3\,b_4\sqrt{3 \Delta^0}\\
\Delta^0&=& -16\,b_2^2\,(b_3^2-4\,b_2b_4)-8\,a_1b_3\,v_{EW}^2(2\,b_3^2-9\,b_2b_4)+27\,a_1^2b_4^2\,v_{EW}\,^4\, .\nonumber
\end{eqnarray}

\begin{figure}[tb]
\centering
\includegraphics[width=0.75\textwidth,clip]{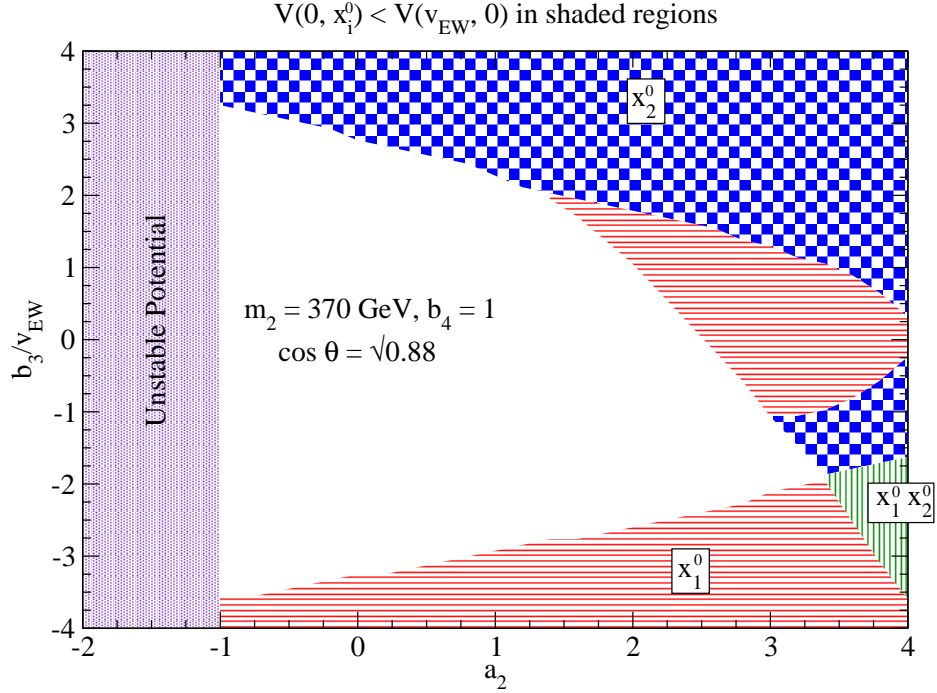}
\caption{Structure of  the 
$v=0$ vacua in the $b_3$ vs. $a_2$ plane for $m_{2}=370$~GeV, $b_4=1$, and $\cos\theta=\sqrt{0.88}$.  The different regions are where the $(v,x)=(v_{EW},0)$ minimum lies below the $v=0$ minima (white region), $(0,x_1^0)$ lies below $(v_{EW},0)$ (red lined),  $(0,x_2^0)$ lies below $(v_{EW},0)$ (blue squares), and both $(0,x_2^0)$ and $(0,x_1^0)$ lie below $(v_{EW},0)$ (green hashed).}
\label{Zero.fig}
\end{figure}

In Fig.~\ref{Zero.fig}, we show the vacuum structure of the $\vev{\phi_0}=0$ minima compared to the $(v,x)=(v_{EW},0)$ minima.  The white region corresponds to where the EWSB minima lies below the  $v=0$ minima, the red lined region to where $(v,x)=(0,x_1^0)$ lies below $(v_{EW},0)$, the blue squares  to where $(0,x_2^0)$ lies below $(v_{EW},0)$, and the green hashed region is where both $(0,x_1^0)$ and $(0,x_2^0)$ lie below $(v_{EW},0)$.  We do not find any region where $V(0,\vev{S}=x_3^0)$ is below the EWSB minima.  Combining the results of Figs.~\ref{NonZero.fig} and \ref{Zero.fig} we can understand the contour in fig.~\ref{fig:EWmin}.